\documentclass[12pt,a4paper,dvips]{article}
\usepackage{epsfig,wrapfig,times,mathptm} 
\renewcommand{\section}[1]{\vspace{6pt} \noindent\mbox{#1} \newline \noindent}
\renewcommand{\subsection}[1]{\vspace{6pt} \noindent\mbox{\underline{#1}} 
\newline \noindent}
\renewcommand{\subsubsection}[1]{\vspace{6pt} \noindent\mbox{\underline{#1}}
\noindent}

\newfont{\sansb}{cmssbx10}
\newfont{\sans}{cmss10}

\begin{document}
{\small OG 4.3.13 \vspace{-24pt}\\}     
{\center \LARGE FIRST RESULTS FROM A SEARCH FOR TEV EMISSION FROM BL LACS
OUT TO Z=0.2
\vspace{6pt}\\}

M. Catanese$^1$, 
P.J. Boyle$^2$, 
J.H. Buckley$^3$, 
A.M. Burdett$^4$, 
J. Buss\'{o}ns Gordo$^2$, 
D.A. Carter-Lewis$^1$, 
M.F. Cawley$^5$, 
D.J. Fegan$^2$, 
J.P. Finley$^6$, 
J.A. Gaidos$^6$, 
A.M. Hillas$^4$, 
F. Krennrich$^1$, 
R.C. Lamb$^7$, 
R.W. Lessard$^6$, 
C. Masterson$^2$, 
J.E. McEnery$^2$, 
G. Mohanty$^1$, 
J. Quinn$^{2,3}$, 
A.J. Rodgers$^4$, 
H.J. Rose$^4$, 
F.W. Samuelson$^1$, 
G.H. Sembroski$^6$, 
R. Srinivasan$^6$, 
T.C. Weekes$^3$, 
and J. Zweerink$^1$ \vspace{6pt}\\
{\it $^1$Dept. of Physics and Astronomy, Iowa State University, Ames, IA 50011, USA\\
$^2$Dept. of Exp. Physics, University College Dublin, Belfield, Dublin 4, Ireland\\
$^3$F. L. Whipple Observatory, Harvard-Smithsonian CfA, P.O. Box 97, Amado, AZ 85645, USA\\
$^4$Department of Physics, University of Leeds, Leeds, LS2 9JT, Yorkshire, England, UK\\
$^5$Physics Department, St. Patrick's College, Maynooth, County Kildare, Ireland\\
$^6$Department of Physics, Purdue University, West Lafayette, IN 47907, USA\\
$^7$Space Radiation Laboratory, California Institute of Technology, Pasadena,
CA 91125, USA  \vspace{-12pt}\\}
{\center ABSTRACT\\}
Two active galactic nuclei have been detected at TeV energies using
the atmospheric \v{C}erenkov imaging technique.  The Whipple
Observatory $\gamma$-ray telescope has been used to observe all the BL
Lacertae objects in the northern hemisphere out to a redshift of 0.1.
We report the tentative detection of VHE emission from a third BL Lac
object, 1ES 2344+514.  
Progress in extending this survey out to z=0.2 will also be
reported.

\setlength{\parindent}{1cm}
\section{INTRODUCTION}
With the detection of very high energy (VHE, $E>250$ GeV) emission
from the two BL Lacertae objects (BL Lacs) , Markarian~421 (Mrk~421)
(Punch et al. 1992) and Mrk~501 (Quinn et al. 1996), we began a survey
of nearby BL Lacs to search for VHE emission.  A collection of
such sources could lead to constraintes on $\gamma$-ray emission
models through investigation into the properties which are important
for VHE emission and also an estimate of the density of extragalactic
background IR light through its effect on the VHE $\gamma$-ray spectra
(Gould and Schr\'{e}der 1967; Stecker, de Jager, \& Salamon 1993).

BL Lacs are blazars, 
the only type of active galactic nucleus (AGN) detected above 100 MeV. 
The electromagnetic spectrum of blazars consists of synchrotron
emission, which spans radio to UV or X-ray wavelengths, produced by
electrons moving within jets oriented at small angles to our line of
sight (Blandford \& K\"{o}nigl 1981), and a high energy part which can
extend to $\gamma$-ray energies.  Models of the high energy emission
fall into three main categories: synchrotron self-Compton emission
(e.g., K\"{o}nigl 1981), inverse-Compton scattering of low energy
photons arising outside the jet (e.g., 
Sikora, Begelman, and Rees 1994),
and pair cascades initiated by protons (Mannheim 1993) or electrons.  
If protons produce the high
energy emission, AGN could contribute significantly to the highest
energy cosmic-ray flux ($E>10^{18}$eV) (Rachen, Stanev, and Biermann
1993).

BL Lacs are particularly promising candidates for VHE emission because
of two aspects of their low energy emission.  First, BL Lacs may have
less $\gamma$-ray absorbing material near the source because they have
weak or no emission lines in their optical spectra (Dermer and
Schlickeiser 1994).  Second, in inverse Compton (IC) models of the
high energy emission, the extension of the synchrotron emission of
X-ray selected BL Lacs (e.g., Mrk~421 and Mrk~501) into the X-ray
waveband implies a higher maximum $\gamma$-ray energy than for
radio-loud BL Lacs (e.g., W~Comae) and flat spectrum
radio quasars (e.g., 3C~279) where the synchrotron emission ends in
the optical to UV range.

Table~\ref{list} lists the objects observed in our BL Lac survey so
far.  We have limited our search to BL Lacs with $z<0.2$ to
reduce the effects of $\gamma$-ray absorption on background IR light.
We observed all BL Lacs with $z<0.1$ visible with the Whipple
Observatory telescope and we are in the process of extending our
survey out to z=0.2.  We applied a two-part approach to the survey.
First, we selected a few promising candidates for long exposures to
search for low-level VHE $\gamma$-ray emission, such as was observed
with Mrk~501 in its initial detection (Quinn et al. 1996).  Second,
the remaining objects were observed for less total time, but the
observations were spread out over a long time period in order to
maximize the chance that we might see an episode of high emission,
such as has been seen in Mrk~421 (Gaidos et al. 1996) and Mrk~501
(Quinn et al. 1996).  This latter approach led to the likely detection
of a third VHE-emitting BL Lac, 1ES 2344+514.

\section{OBSERVATIONS AND ANALYSIS}
The VHE observations reported in this paper were made with the
atmospheric \v{C}erenkov imaging technique (Cawley and Weekes 1995)
using the 10~m optical reflector located at the Whipple Observatory on
Mt.  Hopkins in Arizona (elevation 2.3 km).  The high resolution
camera, consisting of 109 photomultiplier tubes, is mounted in the
focal plane of the reflector and records images of atmospheric
\v{C}erenkov radiation from air showers produced by $\gamma$-rays and
cosmic rays (Cawley et al. 1990).  The energy threshold of the
observations reported here is 350 GeV.

\v{C}erenkov light images are classified according to their angular size
and orientation.  $\gamma$-ray images are typically smaller and more
elliptical than background hadronic images and they are preferentially
oriented toward the source location.  The basic data selection was
based on the Supercuts criteria (Reynolds et al. 1993).  However, some
modifications have been made to account for changes to the telescope
which reduced the detector energy threshold and increased the
background from event triggers caused by fluctuations in night-sky
background light and \v{C}erenkov events caused by single
local muons (see Catanese et al. 1996 for details).

The results reported in this paper use a {\sl Tracking} analysis
wherein events whose orientations do not point toward the object's
direction are used to determine the background level.  A large
collection of non-source data, consisting of off-source observations
and non-detected objects other than BL Lacs, were combined to
estimate the factor which converts the off-source events to a
background estimate.  The count rates are converted to integral fluxes
by expressing them as a multiple of the Crab Nebula count rate and
then multiplying that fraction by the Crab Nebula flux, $I(>350 \ {\rm
GeV}) = 8.7 \times 10^{-11}$ photons cm$^{-2}$ s$^{-1}$ (Hillas et al.
1997).  This procedure assumes that the Crab Nebula VHE $\gamma$-ray
flux is constant, as 7 years of Whipple Observatory data indicate
(Hillas et al. 1997), and that the object's photon spectrum is
identical to that of the Crab Nebula between 0.3 and 10 TeV, $dN/dE
\propto E^{-2.4}$ (Hillas et al.~1997), which may not be the case.
If no significant emission is seen from a candidate source, a 99.9\%
confidence upper limit is calculated using the method of Helene (1983).

\section{RESULTS}
\label{results}
In Table~\ref{list} we present the results of observations for which
the analysis has been completed.  With the exception of 1ES 2344+514,
there is no evidence of emission from any of the objects in this
survey.  In particular, the EGRET sources W~Comae (von Montigny et al.
1995) and BL Lacertae (Catanese et al. 1997a) are not detected despite
long exposures.  Also, only 1ES 2344+514 shows
evidence of short term activity.  
\begin{table}[h]
\vspace{-12pt}
\caption{BL Lac Observations}
\label{list}
\begin{center}
\begin{tabular}{lrccrrrrr} \hline \hline
  &  &  &  &  &  & \multicolumn{1}{c}{Max.} &  &  \\ 
  &  &  &  Observ. & \multicolumn{1}{c}{Exp.} &  & 
\multicolumn{1}{c}{Daily} & \multicolumn{1}{c}{Flux} &  \\ 
\multicolumn{1}{c}{Object} & \multicolumn{1}{c}{z} & 
Type$^a$ & Epoch & \multicolumn{1}{c}{(hrs)} & 
\multicolumn{1}{c}{Excess} & \multicolumn{1}{c}{Exc.} & 
\multicolumn{1}{c}{(Crab)$^b$} & \multicolumn{1}{c}{Flux$^c$} \\ \hline
1ES 2344+514 & 0.044 & X & 1995/96 & 20.5 & 5.3$\sigma$ & 6.0$\sigma$ & 0.16$\pm$0.03 & 1.4$\pm$0.3 \\ 
 &  &  & 20-12-95 & 1.8 & 6.0$\sigma$ &  & 0.63$\pm$0.11 & 5.5$\pm$1.0 \\
 &  &  & 1996 & 32.1 & 1.6$\sigma$ & 2.1$\sigma$ & $<$0.081 & $<$0.70 \\
Markarian 180 & 0.046 & X & 1996 & 20.9 & -0.4$\sigma$ & 0.6$\sigma$ & $<$0.105 & $<$0.91 \\
1ES 1959+650 & 0.048 & X & 1996 & 3.7 & 0.3$\sigma$ & 1.2$\sigma$ & $<$0.128 & $<$1.10 \\
3C 371 & 0.051 & R &  &  &  &  &  &  \\
I Zw 187 & 0.055 & X & 1996 & 2.3 & 0.6$\sigma$ & 1.2$\sigma$ & $<$0.150 & $<$1.30 \\
1ES 2321+419 & 0.059? & X? & 1996 & 6.4 & -1.3$\sigma$ & 0.3$\sigma$ & $<$0.106 & $<$0.92 \\
BL Lacertae$^d$ & 0.069 & R & 1995 & 39.1 & -1.0$\sigma$ & 0.5$\sigma$ & $<$0.06 & $<$0.53 \\
1ES 1741+196 & 0.083 & X & 1996 & 8.8 & -1.7$\sigma$ & 0.6$\sigma$ & $<$0.046 & $<$0.40 \\
W Comae & 0.102 & R & 1996 & 16.6 & -0.4$\sigma$ & 1.4$\sigma$ & $<$0.056 & $<$0.47 \\ 
1ES 0145+138 & 0.125 & X &  &  &  &  &  &  \\
EXO 0706.1+5913 & 0.125 & X &  &  &  &  &  &  \\
1H 1219+301 & 0.130 & X &  &  &  &  &  &  \\
1H 1430+423 & 0.130 & X &  &  &  &  &  &  \\
1ES 0229+200 & 0.139 & X &  &  &  &  &  &  \\
1ES 1255+244 & 0.141 & X &  &  &  &  &  &  \\
1H 0323+022 & 0.147 & X &  &  &  &  &  &  \\
1ES 0927+500 & 0.188 & R &  &  &  &  &  &  \\ \hline \hline
\multicolumn{9}{p{6in}}{$^a$Indicates whether the object 
is radio selected (R) or X-ray selected (X).} \\
\multicolumn{9}{p{6in}}{$^b$Flux,
or upper limit, is expressed in units of the Crab Nebula flux.} \\
\multicolumn{9}{p{6.5in}}{$^c$Integral fluxes or upper limits 
are quoted above 350 GeV in units of $10^{-11}$cm$^{-2}$s$^{-1}$.  
Flux upper limits are at the 99.9\% confidence level.}\\
\multicolumn{9}{p{6in}}{$^d$Results from Catanese et al.
1997a.} \\
\end{tabular}
\end{center}
\end{table}

Most of the excess from 1ES 2344+514 during 1995 comes from an
apparent flare on 1995 December 20 (see Figure~1).  We find a
non-statistically significant excess from this object in 1996 which
could simply mean that the average flux level dropped below the
telescope sensitivity limit, as occasionally happens with Mrk 421
(Buckley et al. 1996).  We currently consider the detection tentative
because we see no evidence for a consistent signal nor is there
independent confirmation of a high state for this object (e.g., from
X-ray observations) during this period.

\begin{wrapfigure}{r}{3.8in}
\label{3elc}
\vspace*{-0.2in}
\epsfig{file=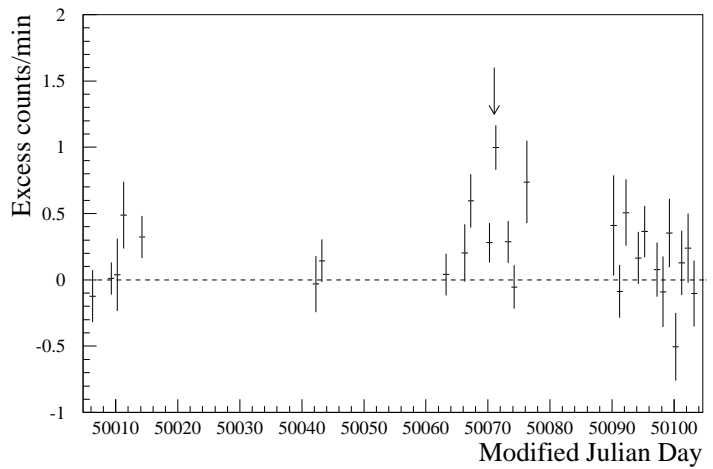,height=2.5in,angle=0.}
\vspace*{-0.15in}
\caption{\it The light curve for $\gamma$-ray observations of 1ES
2344+514 between Oct. 1995 and Jan. 1996.  The flare on 20-12-1995 is
indicated by the arrow.}
\end{wrapfigure}

\section{CONCLUSIONS}
Both Mrk 421 and Mrk 501 have energy spectra, expressed as $\nu
F_\nu$, with peaks of comparable amplitude at the X-ray and
$\gamma$-ray energies (Buckley et al. 1996; Catanese et al.
1997b).  If the X-ray selected BL Lacs in this sample are similar to
Mrk~421 and Mrk~501, an upper limit below the X-ray power output would
indicate some reduction of the VHE flux at the source or en-route to
Earth.  Our current limits are not below the lowest measured X-ray
fluxes for these objects.  Further observations will be needed to
clearly establish whether these objects emit VHE $\gamma$-rays at the
expected levels.

For the two EGRET sources in this survey, W~Comae and BL Lacertae, the
Whipple upper limits are well below the power output at EGRET energies
(see Figure~2 and Catanese et al. 1997a).  This is consistent with IC
emission models because these objects have synchrotron breaks in the
optical-UV waveband and so would have lower maximum $\gamma$-ray
energies.  However, the VHE observations are not contemporaneous, so
it may be that the $\gamma$-ray emission was in a low state when these
objects were observed at the Whipple Observatory.  Also, the VHE
$\gamma$-ray emission from W~Comae may have been attenuated
significantly by the IR background because its redshift is 0.102.

\begin{wrapfigure}{r}{3.5in}
\label{nufnu}
\vspace*{-0.2in}
\epsfig{file=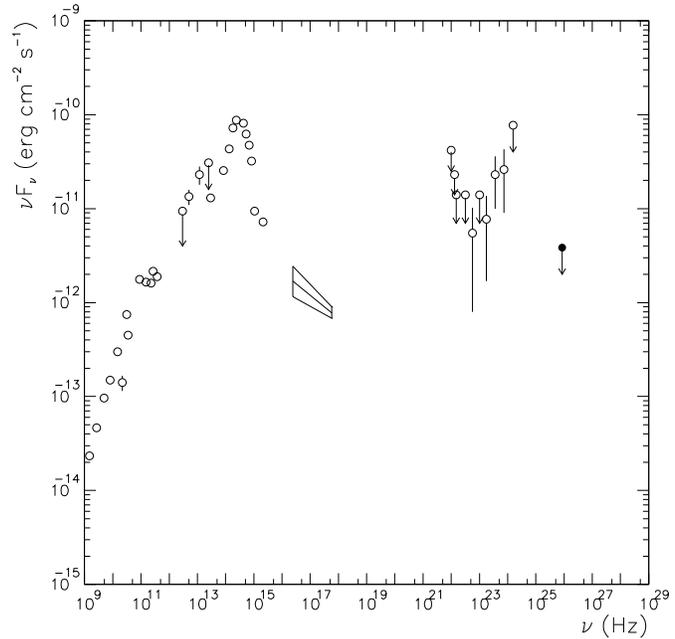,height=3.4in,angle=0.}
\vspace*{-0.3in}
\caption{\it Spectral energy distribution of W~Comae.  Shown are 
the VHE upper limit (filled circle) and 
archival data (open circles) (von Montigny et al. 1995; Lamer, Brunner and Staubert 
1996; Edelson et al. 1992; Cruz-Gonzalez and Huchra 1984; 
Impey and Neugebauer 1988; Gear et al. 1994; Owen, Spangler, and Cotton 1980).}
\end{wrapfigure}

Finally, the detection of 1ES~2344+514, if confirmed, would be
consistent with the other detected VHE blazar sources: it is a nearby
X-ray selected BL Lac (the third closest known) whose
$\gamma$-ray emission is variable.

\section{ACKNOWLEDGEMENTS}
This research is supported by grants from the U. S. Department of
Energy and NASA, by PPARC in the UK and by Forbairt in Ireland.

\section{REFERENCES}
\setlength{\parindent}{-5mm}
\begin{list}{}{\topsep 0pt \partopsep 0pt \itemsep 0pt \leftmargin 5mm
\parsep 0pt \itemindent -5mm}
\vspace{-15pt}

\item Blandford, R. D., \& K\"{o}nigl, A., {\it ApJ}, 232, 34 (1979).

\item Buckley, J. H., et al., {\it ApJ}, 472, L9 (1996).

\item Catanese, M., et al., in {\it Towards a Major Atmospheric Detector-IV},
ed. M. Cresti, 335 (1996).

\item Catanese, M., et al., {\it ApJ}, 480, 562 (1997a).

\item Catanese, M., et al., {\it ApJ}, submitted (1997b).

\item Cawley, M. F., \& Weekes, T. C., {\it Exp. Astron.}, 6, 7 (1995).

\item Cawley, M. F., et al., {\it Exp. Astron.}, 1, 173 (1990).

\item Cruz-Gonzalez, I., and Huchra, J. P., {\it AJ}, 89, 441 (1984).

\item Dermer, C. D., \& Schlickeiser, R., {\it ApJS}, 90, 945 (1994).


\item Edelson, R., et al., {\it ApJS}, 83, 1 (1992).


\item Gaidos, J. A., et al., {\it Nature}, 383, 319 (1996).

\item Gear, W. K., et al., {\it MNRAS}, 267, 167 (1994).

\item Gould, J. R., \& Schr\'{e}der, G. P., {\it Phys. Rev.}, 155, 1408 (1967).

\item Helene, O., {\it Nucl. Instr. Meth.}, 212, 319 (1983).

\item Hillas, A. M., et al., in preparation (1997).

\item Impey, C. D., and Neugebauer, G., {\it AJ}, 95, 307 (1988).

\item K\"{o}nigl, A., {\it ApJ}, 243, 700 (1981).

\item Lamer, G., Brunner, H., and Staubert, R., {\it astro-ph/9603030} (1996).

\item Mannheim, K., {\it A\&A}, 269, 67 (1993).

\item Owen, F. N., Spangler, S. R., and Cotton, W. D., {\it AJ}, 84, 351 (1980).



\item Punch, M., et al., {\it Nature}, 358, 477 (1992).

\item Quinn, J. et al., {\it ApJ}, 456, L83 (1996).

\item Rachen, J. P., Stanev, T., and Biermann, P. L., {\it A\&A}, 
273, 377 (1993).

\item Reynolds, P. T., et al., {\it ApJ}, 404, 206 (1993).

\item Sikora, M., Begelman, M. C., \& Rees, M. J., {\it ApJ}, 421, 153 (1994).

\item Stecker, F. W., de Jager, O. C., \& Salamon, M., {\it ApJ}, 
415, L71 (1993).


\item von Montigny, C., et al., {\it ApJ}, 440, 525 (1995).


\end{list}

\end{document}